\documentclass[showpacs,pre,preprintnumbers,amsmath,amssymb,10pt]{revtex4}
\usepackage{amsmath}
\usepackage{amsfonts}
\usepackage{amssymb}
\usepackage{graphicx}
\usepackage{times}
\usepackage{subfigure}
\newcommand{\refE}[1]{Equation~(\ref{#1})}

\newcommand{\refe}[1]{Eqn.~(\ref{#1})}

\newcommand{\reff}[1]{Fig.\,\ref{#1}}

\newcommand{\bfX}{\mathbf{X}}
\newcommand{\bfZ}{\boldsymbol{Z}}

\begin{document}
\author{Sergey Plyasunov}
\affiliation{
 Physics Department, UC Berkeley,
Physical Biosciences Division, E.O. Lawrence Berkeley National
Laboratory, 1 Cyclotron Road, Berkeley, CA 94720, Email:
teleserg@uclink.berkeley.edu}

\title{\bf
  Averaging methods for stochastic dynamics of complex reaction networks:\\
  description of multi-scale couplings
}

\fontsize{11.0pt}{11.0pt}\selectfont 
\date{\today}
\begin{abstract}
This paper is concerned with  classes of models of stochastic reaction
 dynamics with time-scales separation. We demonstrate that  the existence of the time-scale separation  naturally leads to the application of the
 averaging principle and elimination of degrees of freedom via the  renormalization of transition rates of slow reactions.
 The method suggested in this work   is more general than  other
approaches presented previously: it is not limited to a particular type of  stochastic processes  and can be applied to
 different types of processes  describing  fast dynamics, and also provides  crossover to the case when separation of
 time scales is not well pronounced. We derive a family of exact fluctuation-dissipation
relations which establish the connection between effective rates and the statistics of the reaction events in  fast reaction channels.   An illustration of the technique is  provided. Examples show that renormalized transition rates  exhibit in general non-exponential relaxation behavior 
 with a  broad range  of possible scenarios. 
\end{abstract}

\maketitle

\section{\large Introduction\label{sec:intro}}
\par
Chemical reaction networks are systems  of 
molecular species of different types interacting with each other by means of multiple reactions 
\cite{FeinbergChemNetworks95}.  In classical chemical systems, the  volume of the reactor and
population numbers of species of each types are usually large giving  the accurate  description  of the system in terms of the
concentrations.  Reactors with complex chemistry give rise to complicated systems of
 nonlinear equations for the concentrations of chemical species that do not lend themselves to analytic solution. 
Dynamics of these quantities can be modeled via  sets of ordinary differential equations (ODEs) which are powerful tools for predicting 
the dynamical behavior of macroscopic  chemical mixtures.
 \par
There is a recent  renewal of interest in  stochastic modeling of chemical systems which came
 with the recent realization of importance of noise in cellular information processing.
At the level of a single cell,  number of molecules involved in some processes can be very
small and concentrations are described as nano-molar \cite{McAdamsArkin97,Oudenaarden02}.
In addition to that, different processes are also characterized by
significantly different times scales \cite{ErbanOthmer2005}. 
\par
 Presence of this  time-scale separation and highly different copy numbers of molecular
species   usually complicates the study of biological processes with
computer simulations.
 There is an obvious need for computationally tractable
stochastic models on a macro-scale that can provide insights into joint, qualitative,
 effects arising  from interaction of several sub-networks.   
 In deterministic systems of ordinary differential equations, time-scale separation is usually related to
the concept of stiffness.  It is obviously hard to define the same
concept in case of the stochastic systems \cite{RathinamGillespie2003}.
 \par
 In spite of these obvious complications some progress has been made in
modeling of biochemical networks which express the separation of
time-scales. One difficulty  is heterogeneity of simulation
techniques used for simulation of ODEs/SDEs and stochastic
simulation algorithm. One strategy exploited in the literature \cite{GillespieTau,RathinamGillespie2003}
is based on grouping together of reaction events taking place in a single reaction channel in a  fast succession
 and applying  diffusion approximation \cite{EthierKurtzBook}. In \cite{CVRaoArkin2003} Rao et.al. 
  discuss a computational  approach for performing    elimination of the  fast species based on rapid equilibrium  in the limit of the infinite time-scale separation. This method was termed  quasi-steady state approximation (QSSA).
A somewhat similar approach  is  taken  in  \cite{HaseltineRowlings2002}.
Formally, this method stems from the classical deterministic QSSA
applied to the chemical master equation itself rather then to the
(stochastic) differential equation underlying the dynamics of the
state vector (numbers of molecular species). The method developed by Cao et al. in
 \cite{CaoGillespiePetzold2005}  can be viewed as  generalization of  approach of Rao et.al. 
\cite{CVRaoArkin2003}   but still have the limitations of being derived through the application of
deterministic techniques and assumptions to the chemical master
equation. It also assumes that averaging procedure can be done by
solving the system of algebraic equations for the expectations of
the fast variables given slow, termed  in \cite{CaoGillespiePetzold2005} as a virtual fast process.
 We note here that studies of stochastic dynamics of diffusion-type processes evolving on different time scales
were pioneered by Bogolubov, Khasminski and Freidlin and we refer the reader to monographs
 \cite{FreidlinWentzelBook,GihmanSkorohodSDE,SkorohodAsymptoticSDE}.

\par
 This paper  has two purposes. First, we present the formulation of
 stochastic  reaction dynamics of reaction network consisting of two subnetworks.
  Compared to many previous results,  where   usual  description of stochastic  reaction dynamics
follows the approach based on chemical master equation (CME), current publication  
follows the path-sampling approach and represents the dynamics as a jump-type stochastic differential equations (SDEs).
\par
Second purpose is to provide rigorous procedure for the
 renormalization of the transition rates of  slow reactions in the presence of fast
 ones. Following the picture of the stochastic dynamics  developed in the first part of this  paper,
 we outline the main guidelines for use of stochastic averaging principle including error control analysis. 
Despite of the recent rebirth of interest to the method of stochastic averaging in applications to stochastic chemical kinetics, very few examples deal with situations when this procedure might break down.  We demonstrate 
here, in  a constructive way, how to perform the  averaging over fast reaction events and how to  obtain 
 the effective slow-scale transition rate. 
 \par
Organization of this  paper is  as follows. In the next section we
discuss the  general  probabilistic framework for stochastic dynamics of
reaction  networks and introduce  a  scheme for the partition of species
and  reactions. In  Section \ref{Sec::time_scales} we investigate the
 consequences of  possible  time-scales separation   and
present a  procedure based on renormalization of transition rates.
We also  put emphasis on  error analysis, outlining main sources of the numerical error
 on  different steps of the procedure.  Our paper will end with discussion of  examples.
\par
\section{\large Network partitioning }
\par
We begin our discussion with a general set-up, introducing basic concepts and notation. 
\par 
Assume that a well mixed, isothermal
system has $S$ different molecular species indexed by $i=1\ldots S$ and there are
$R$ reaction channels, index by $r=1\ldots R$, transforming the molecualr composition of these species.
 For the basic notation and examples  we direct  reader to \cite{DTGillespieBook,GillespieSSA}. 
State  vector of the system can be represented as following:
\begin{gather}\label{}
(\mathbf{X},\boldsymbol{Z})
\end{gather}
where fist part of the state vector  ${X}_{i}\in\mathbb{Z}_{+},i=1\dots
S_X$ represents {\em main} species  while the second part ${Z}_{i}\in\mathbb{Z}_{+},i=1\dots
S_z$ represents {\em intermidiate} species $Z_i,i=1\dots S_Z$. Total number of all types of species:  $S_X+S_{Z}=S$.
Vectors $\boldsymbol{\nu}^{X}_r$, $\boldsymbol{\nu}^{Z}_r$ and $\boldsymbol{\nu}^{XZ}_r$
are  stoichiometric changes of components ${\bf X}$ and
$\boldsymbol{Z}$ if reaction event $r$ takes place. We will not make any assumptions about 
actual number of molecular species of each type, i.e. we will not assume low or large copy numbers.
\par
We assume, however,  that there are three subsets of reactions in the system: \par
(i) reactions which transform only species $\mathbf{X}$
(we denote this subset  $\mathcal{R}_1$),\par
(ii) reactions which transform only species $\boldsymbol{Z}$ (subset $\mathcal{R}_2$) 
\par 
(iii) "linker" reactions which mix species $\mathbf{X}$ and $\mathbf{Z}$ (subset $\mathcal{R}_3$).
\par
Each reaction channel can be specified  by the
transition rates $a_r$ (a positive function) which describes the probability
$a_rdt$ of reaction event to take place in the interval of time
$dt$. Transition rate $a_r$ can be further specified
 as positive functions  of $\mathbf{X}$, $\bfZ$, or,in general,  on both components $\mathbf{X}$ and $\boldsymbol{Z}$. Based on the definition of subsets
 $\mathcal{R}_{1,2,3}$ we have:
\begin{subequations}\label{Eq::transit_rates}
 \begin{gather}
  a_r(\mathbf{X}),\quad r\in \mathcal{R}_1\\
  a_r(\boldsymbol{Z}),\quad r\in \mathcal{R}_2\\
  a_r(\mathbf{X},\boldsymbol{Z}),\quad r\in \mathcal{R}_3
 \end{gather}
\end{subequations}
We do not assume specific dependence of $a_r(\cdot)$ on the state variables $\bfX$ and $\bfZ$ but usually,
in the framework of mass action kinetics, it is a product 
of kinetic rate  $k_r$  and function $h_r(\cdot)$  which  represents the number of reactive configurations available at a given state $\mathbf{X},\boldsymbol{Z}$ \cite{DTGillespieBook}.
\par
There exist different methods to characterize the stochastic chemical dynamics. One of the most popular approach 
is to provide an equation for the joint probability  density $p_t(\mathbf{X},\boldsymbol{Z})$, which gives all information
about instantaneous state of the system at generic moment of time $t$. Such equation is known as chemical master equation
(CME) \cite{vanKampen92,DTGillespieBook} and it has been intensively described and utilized in recent literature 
\cite{CaoGillespiePetzold2005,CVRaoArkin2003,HaseltineRowlings2002}. But even if we can 
obtain \cite{SamoilovCME} the solution of CME, which is usually  a very hard problem even for simple chemical networks, this approach still have certain limitations, coming from instantaneous description 
provided by the density $p_t(\cdot)$.
\par
To describe the stochastic dynamics of the chemical network one can  introduce  the set of independent
point processes  $N_r(t), N_r(0)=0$ representing the numbers of reaction events
which took place in channels $r\in\mathcal{R}$ up to time $t$ and use the mass balance relations:
\begin{subequations}
\label{Eq::stoch_dynamics}
\begin{gather}
\mathbf{X}_{t}=\mathbf{X}(0)+\sum_{r\in\mathcal{R}_1}\boldsymbol{\nu}^{X}_{r}N_r(t)+\sum_{r\in\mathcal{R}_3}\boldsymbol{\nu}^{XZ}_{r}N_r(t),\\
\boldsymbol{Z}_{t}=\boldsymbol{Z}(0)+\sum_{r\in\mathcal{R}_2}\boldsymbol{\nu}^{Z}_{r}N_r(t)+
\sum_{r\in\mathcal{R}_3}\boldsymbol{\nu}^{XZ}_{r}N_r(t),
\end{gather}
\end{subequations}
where vectors $\boldsymbol{\nu}^{Z}_{r},\boldsymbol{\nu}^{X}_{r}$ and $\boldsymbol{\nu}^{XZ}_{r}$ describe 
the composition change of the system due to the reaction event in the channel $r$.
 Average number of reaction events in each reaction channel $r\in \mathcal{R}_{1,2,3}$ during the small time interval
 $[t,t+\delta t)$ are proportional to the transition rates  (\ref{Eq::transit_rates}):
\begin{gather}
\mathbb{E}(N_r(t+\delta t)-N_r(t)|\mathbf{X}_{t},\boldsymbol{Z}_{t})=a_r(\mathbf{X}_{t},\boldsymbol{Z}_{t})\delta t
+O(\delta t^2)
\end{gather}
\par
Processes $N_r(t)$ can be considered as  time-changed,  unit-rate independent Poisson processes 
$\Pi_r(t)$  \cite{EthierKurtzBook}:
\begin{subequations}
\begin{gather}
\label{Eq::Poisson_process}
N_r(t)=\Pi_r(\int_0^ta_r({\mathbf{X}}_{t'},\boldsymbol{Z}_{t'}) dt')
\end{gather}
Thus, the large class of discrete event systems  with totally inaccessible event times 
can be viewed as a standard {\em Poisson process} with appropriate change of the time scale:
\begin{gather}
t \mapsto \int_0^t a_r({\mathbf{X}}_{t'},\boldsymbol{Z}_{t'}) dt'
\end{gather}
\end{subequations}
 The time change generates path-dependent or self-affecting point processes whose dynamics depend on the 
information generated by the arrivals of the process $(\mathbf{X}_{t},\boldsymbol{Z}_{t})$ . 
It is important to take into account that the stochastic differential equation does not only introduce the probability distribution for the pair
 $(\mathbf{X},\boldsymbol{Z})$ but also generates a measure on the paths, 
which contains much more information. For almost any realization of the set of $1\ldots R$ standard Poisson processes,
 $\Pi_r(t,\omega)$, parametrized by the element $\omega$ of event space 
 \cite{KaratzasShreveBook,EthierKurtzBook} and any deterministic initial condition the
 solution $(\mathbf{X}(t,\omega),\boldsymbol{Z}(t,\omega))$
is a step-wise stochastic process.\par 
 Note also, that dynamics  of each component $\mathbf{X}$ or $\boldsymbol{Z}$ is
  non-Markovian if considered separately but the dynamics of the
  pair  $(\mathbf{X},\boldsymbol{Z})$ is Markovian.\par
 So far we have introduced only the basic notation: quite generic system of SDEs
 given by (\ref{Eq::stoch_dynamics}) outlined in this section have not invoked any assumptions
on particular relations between different transition rates  $a_r$
and was totally based on prior information about existence of two groups of species, i.e. $X_i$ and $Z_i$ which uniquely identified the partition of the reactions into the subsets $\mathcal{R}_1,\mathcal{R}_2$ and $\mathcal{R}_3$ .
\par
In the next section we consider the particular implication of
time-scale separation including the extensions of the stochastic
averaging principle and diffusion approximation.
\par
\section{\large Separation of time-scales and Elimination of Fast Stochastic Variables.
\label{Sec::time_scales}}
\par
In many situations, dynamics of main species $\bfX$ is propagated via large number of fast transitions which transform mainly intermediate species $\bfZ$.
 One usually desires to construct an approximate, time coarse-grained model, which involve only main species.
   It is important that  approximate problem describes the dynamics of the system on a large time scale and thus
is more advantageous for performing  simulations without significant sacrifice in accuracy.
  This section deals with substitution of the original problem with approximate one and
demonstrates the form convergence of the approximation under
certain assumptions. 
\par We assume that at certain region of state space the 
following assumption can be
made about transition rates $a_r(\cdot)$:
 \begin{gather}
 \sum_{r\in\mathcal{R}_1\cup\mathcal{R}_3}a_r\propto O(1) {\rm ~while~~}
\sum_{r\in \mathcal{R}_2}a_r\propto O({\epsilon}^{-1})
 \end{gather}
where  separation of the time-scales is introduced via the small
parameter $\epsilon\ll 1$.
Problems of this type are  challenge for direct application of  Stochastic Simulation Algorithm (SSA)
 \cite{BortzKalosLebowitz75,GillespieSSA}  because  they  will require the
time steps of the order $O(\epsilon)$ with a total computational cost of order
$\epsilon^{-1}$. If we want to advance through the time interval $[0,t],~t\sim O(1)$  most of the simulation
 time will be spent on simulation of reaction events with the high intensity 
 ($\sum_{r\in \mathcal{R}_2}a_r\propto O({\epsilon}^{-1})$).
We would like to find an effective transition rates $\bar{a}_r(\cdot)$ for the "linker" reactions (subset
$\mathcal{R}_{3}$), which describe the transition events of the slow reactions "coarse-grained" over
the possible events corresponding to the reaction events in subset
$\mathcal{R}_2$.
\par It is instructive to consider  a simple reaction scheme involving three species $\mathsf{X}_1,\mathsf{Z}_{1,2}$ similar to one considered in \cite{CaoGillespiePetzold2005}:
\begin{gather}
\mathsf{Z}_1\overset{k_2\epsilon^{-1}}{\underset{k_3\epsilon^{-1}}
{\rightleftharpoons}} \mathsf{Z}_2 \overset{k_1}{\rightarrow}\mathsf{X}_1
\end{gather}
where rates $k_{1,2}\propto \epsilon^{-1}$ are parametrized by small $\epsilon$ and  $k_3\propto O(1)$.
In this case reactions $\mathsf{Z}_1
{\rightleftharpoons} \mathsf{Z}_2$ forms the subset $\mathcal{R}_2$ 
while reaction $\mathsf{Z}_2 {\rightarrow}\mathsf{X}_1$ 
corresponds to the subset $\mathcal{R}_3$ and subset $\mathcal{R}_1 $ is empty,i.e $\mathcal{R}_1=\{\emptyset\}$.
Then systems of equations for components $(X_1,Z_1,Z_2)$ is the following one:
\begin{subequations}
\begin{gather}
Z_{1t}=Z_{10}-N_2(t)+N_3(t),\\
Z_{2t}=Z_{20}+N_2(t)-N_3(t)-N_1(t),\\
X_{1t}=X_1(0)+N_1(t)
\end{gather}
\end{subequations}

Presence of the scaling factor $\epsilon^{-1}$ in reaction constants $k_{1,2}\epsilon^{-1}$ allows us to consider family of solutions 
parameterized by $\epsilon$. We expect  $Z_{1,2}$ to follow adiabatically the $X_{1t}$. To make that apparent,
 one can apply the functional law of large numbers to the processes $N_{2,3}(t)$ in  time interval $[0,t]$  
(see \refe{Eq::Poisson_process}):
\begin{gather}
N_2(t)-N_3(t){\rightarrow} \frac{1}{\epsilon}\left(\int_0^t k_2Z_{1t'}dt'-\int_0^tk_3Z_{2t'}dt'\right)+\\+
\frac{1}{\sqrt{\epsilon}}\left(W_2(\int_0^t k_2Z_{1t'}dt')-W_3(\int_0^tk_3Z_{2t'}dt')\right),~\epsilon\to 0
\end{gather}
where $W_{2,3}(\cdot)$ are two independent Wiener processes \cite{EthierKurtzBook}. 
Since  parameter $\epsilon^{-1}$  is large, we can  conclude that difference 
\begin{gather}\nonumber
\left|\int_0^t k_2Z_{1s}ds-\int_0^t k_3Z_{2s}ds\right|
\end{gather}
 also converges  to zero for times $t\leq\ \epsilon/(k_2+k_3) $ in the limit of small $\epsilon$, and we can conclude that:
\begin{gather}
\sup_{0\leq t'\leq t}|k_2Z_{1t'} - k_3Z_{2t'}|\to 0
\end{gather}
This means that    variables  $Z_{1t}$  and $Z_{2t}$ reach a stationary binomial distribution:
\begin{gather}
\pi^{\epsilon\to 0}(Z_1,Z_2|X_1)\propto \alpha^{Z_1}(1-\alpha)^{Z_2},\\
\quad
Z_0=Z_1(0)+Z_2(0)=Z_1+Z_2,~
\alpha= \frac{k_2Z_0}{k_2+k_3}
\end{gather}
on the time scale $t\propto O(\epsilon )$ while sum $Z_{1t}+Z_{2t}$ changes on the much larger time-scale $t\geq O(1)$:
\begin{subequations}
\begin{gather}
Z_{1t}+Z_{2t}\approx Z_1(0)+Z_2(0)-N_1(t),\\
X_{1t}\approx N_1(t)
\end{gather}
\end{subequations}
\par
By exploiting the separation of time-scales  
using the stationary distribution $\pi^{\epsilon}(Z_1,Z_2|X_1)$ one can replace dynamical quantities
$f(Z_{1t},Z_{2t},X_{1t})$ averaged on the time interval $[0,t],\quad ~ 
\frac{\epsilon}{k_2+k_3}\ll t <\frac{1}{k_1}$ with their conditional averages:
\begin{gather}
\label{Eq::averaging_sigma} f(Z_{1t},Z_{2t},X_{1t})\approx
\frac{1}{t }\int_{0}^{t }f(Z_{1t'},Z_{2t'},X_{1t'})dt'\approx\\\approx 
\bar{f}(X_{1t})=
\sum_{Z_1,Z_2} f(Z_1,Z_2,X_{1t})\pi^{\epsilon}(Z_1,Z_2|X_{1t})
\end{gather}
and {\em eliminate } fast variables $Z_{1,2}$ from the description even though the total number of molecules 
$Z_1+Z_2$ may be not  a large quantity. Thus, taking $f(\cdot)$ to be the "linker"
 transition rates $a_r(\bfX,\bfZ),~r\in\mathcal{R}_3$ one obtains averaged transition rates $\bar{a}_r(\bfX)$ which now depend only on the slow variable $\bfX$.
 Results of the large deviation theory \cite{FreidlinWentzelBook} demonstrate weak convergence bounds of the original problem with small but non-zero $\epsilon$ to the solution of the averaged system. But as we mentioned it before, one of the goals 
of this publication is to analyze and  extend averaging process to the situation when $\epsilon$ may be  small, but 
not 'infinitesimally' small. In the  next section \ref{sec::renormalize} we will try to answer this question.

\par
\subsection{Renormalization of fast fluctuating reaction rates and reduced evolution equations\label{sec::renormalize}}
\par
Recall that transition rates $a_r(\cdot)$ of a jump Markovian process can
be used to describe distributions of the waiting times of the
reaction events via the survival probability of a given state $(\mathbf{X},\boldsymbol{Z})$ has an exponential form 
$S(t)=e^{-\sum_{r=1}^Ra_r(\mathbf{X},\boldsymbol{Z})t}$ and describes  probability that no reaction
event take place in any of $1\ldots R$ reaction channels in time interval $[0,t]$ \cite{GihmanSkorohod_volII}. 
\par  
Consider the first jump time of a particular
reaction $r$ in the subset of the "linker" reactions, $\tau_{r,3}$
 and first jump times of any reaction in the subset of the fast reactions which we will denote
$\tau_{r,2}$ . Reaction in the group $\mathcal{R}_3$ have both types of chemical species ($\mathbf{X}$ and
$\boldsymbol{Z}$) as their substrates, that means that reaction rates in this subset are fluctuate 
with fast variables  $\boldsymbol{Z}$. If system is originally prepared  at the state
$(\mathbf{X}_0,\boldsymbol{Z}_0)$ at $t=0$ then at any moment
of time $t>0$ one is interested in finding  the probabilities
of events $\{\tau_{r,3}>t\}$ and $\{\tau_{r,2}<t\}$. In other words one has to find an averaged survival probabilities:
\begin{gather}
\label{Eq::survival_prob}
S_{r}(t|\bfX)=P(\{\tau_{r,3}>t\})=\left\langle
\exp(-\int_0^t~a_r(\mathbf{X}_0,\boldsymbol{Z}^x_{t'})dt')
\right\rangle_{Z},
\quad r\in\mathcal{R}_{3}
\end{gather}
Average $\langle \ldots\rangle_{Z}$
stands for the average over the possible trajectories of the stochastic process 
$\boldsymbol{Z}^x([0,t]),~\boldsymbol{Z}^x_0=\boldsymbol{Z}_0$ at {\em fixed} $X$ 
which depends on $\bfX$ as on parameter \cite{FreidlinWentzelBook}. 
 \par
 Probabilities (\ref{Eq::survival_prob}) can be used to introduce time-dependent
transition rates $\bar{a}_r(\mathbf{X},t)$ which effectively
describe the dynamics for reactions in the groups
$\mathcal{R}_{3}$. Taking the logarithm of the averaged survival
probabilities (\ref{Eq::survival_prob}) we obtain:
\begin{subequations}
\begin{gather}\label{Eq::renorm_rates}
S_{r}(t|\bfX)=\exp(-\int_0^tdt'~\bar{a}_r(\mathbf{X},t')),\\
\bar{a}_r(t,\mathbf{X})=-\frac{\partial}{\partial
t}\ln\left\langle\exp(-\int_0^tdt'~a_r(\mathbf{X}_0,\boldsymbol{Z}_{t'})\right\rangle_{Z}
\end{gather}
\end{subequations}
Equations \ref{Eq::renorm_rates} constitute one of the main results of the paper.
In the field of chemical kinetics a similar methodology is known
under the label  of the "rate dependent processes with dynamical
disorder" \cite{BurlatskyOshaninMogutov90,Zwanzig92,AgmonHopfield83,VladRossMackey96,ChandlerGehlenMarchi94,WangWolynes94} where it describes
the influence of the non-equilibrium environmental degrees of freedom on transport and kinetic properties. Similar
approach was used to describe quantum dynamics in fluctuating environment \cite{GoychukPRE2004}.
  Using the procedure of the cumulant expansion \cite{vanKampen92,RKubo69} we can obtain
the following interrelationship between  $\bar{a}_r$ and the
multi-point  cumulants $C_{r}^{(m)}(t_1,\ldots,t_m|\mathbf{X})$ of the functions
$a_r(\mathbf{X},\boldsymbol{Z}^x_{\cdot})$, taken at different temporal points
$t_1,\ldots, t_m$:

\begin{gather}\label{Eq::cumulant_expansion}
S_{r}(t|\bfX)=\exp\left[\sum_{m\geq
0}\frac{(-1)^m}{m!}\int_0^tdt_1\ldots\int_0^tdt_mC_{r}^{(m)}(t_1,\ldots,t_m|\mathbf{X})\right],\\
\bar{a}_r(t,\mathbf{X})=\left\langle a_r(\mathbf{X},\boldsymbol{Z}_{t})\right\rangle_{{Z}}+
\sum_{m\geq~2}\frac{(-1)^{m-1}}{m!}\int_0^tdt_1\ldots\int_0^t~dt_m~C_{r}^{(m)}(t_1,\ldots,t_m|\mathbf{X})
\end{gather}
Renormalized transition rates $\bar{a}_r(t,\mathbf{X})$ provide  so-called
semi-Markov  approximation \cite{GihmanSkorohod_volII,vanKampen92}. Term
"semi-Markov" generally describes non-Markov processes since the
statistical properties of the waiting times can not be provided only by average rate of the process
 but   all the multi-time joint probability distributions for the
considered process must be considered. Note that in our case effective rate $\bar{a}_r$ depend on the statistics of 
fluctuations of fast variables $\boldsymbol{Z}$ through the  cumulants $C_{r}^{(m)}(t_1,\ldots,t_m|\mathbf{X})$.
\par
 Taking a leading term at $\epsilon\to 0$, which  sometimes called {\em Markovian limit},
 we formally  arrive to the results of the QSS Approximation \cite{CVRaoArkin2003}:
\begin{gather}\label{Eq::markovian_limit}
\bar{a}_r(\mathbf{X},t)=C^{(1)}_r(t|\mathbf{X})=\lim_{\epsilon\to 0}
\sum_{Z}
a_r(\mathbf{X},\boldsymbol{Z})\pi_{X}^{\epsilon}(\boldsymbol{Z})
\end{gather}
where average is taken over the {\em invariant measure} $\pi^{\epsilon}(\boldsymbol{Z}|\mathbf{X})$
of the fast process $\boldsymbol{Z}^x_{t}$ at {\em fixed } ${\bf X}$. Note that at this level 
$\bar{a}_r$ does  not depend
on time  and correspond to the single exponential form of the survival probability.
This level of approximation corresponds to the assumption that at
fixed $\mathbf{X}$ all state space of $\boldsymbol{Z}$ is
totally accessible, i.e. {\em ergodic} \cite{FreidlinWentzelBook} and for any function
$f(\cdot):\mathbb{Z}^{n_Z}\to \mathbb{R}$:
\begin{gather}
\bar{f}(\mathbf{X})=\lim_{t\to\infty}t^{-1}\int_0^t
f(\mathbf{X},\boldsymbol{Z}^x_{s})ds= \lim_{\epsilon\to
0}\sum_{Z}f(\mathbf{X},\boldsymbol{Z}^x)\pi^\epsilon(\boldsymbol{Z}^x|\mathbf{X})
\end{gather}
\par
There is a general Jensen inequality , which gives the relationship
between the mean value of a convex function of a random variable an
the value of this function when its argument equals the mean value
of the random variable. According to this inequality:
\begin{gather}
S_{r}(t|\bfX)\geq \exp\left(-\int_0^tdt' C^{(1)}_r(t'|\mathbf{X})\right)
\end{gather}
Application of this inequality leads to the important conclusion that  mean field rate (\ref{Eq::markovian_limit})
 is larger then the rate given by  (\ref{Eq::cumulant_expansion}). The exponential and non-exponential structure of the averaged survival probability is governed by the hierarchy of the time scales of the dynamics of $\boldsymbol{Z}_{t}$ at different values of $\mathbf{X}$.
If dynamics of $\boldsymbol{Z}$ is  complicated and exhibit metastability at some values of ${\bf X}$
then Markovian approximation \ref{Eq::markovian_limit} is no longer holds and additional corrections
 corresponding to the high order cumulants must be taken into  consideration. Correction to the Markovian approximation
based on the second order cumulants is:
\begin{subequations}
\begin{gather}\label{Eq::second_order_correction}
\Delta \bar{a}_r(t,\mathbf{X})\cong -\int_0^t~dt'~C^{(2)}_r(t,t'|\mathbf{X}),\\
C^{(2)}_r(t,t'|\mathbf{X})\equiv
\left\langle 
a_r(\mathbf{X},\boldsymbol{Z}^x_{t})a_r(\mathbf{X},\boldsymbol{Z}^x_{t'})
\right\rangle_{Z}-
\left\langle a_r(\mathbf{X},\boldsymbol{Z}^x_{t})\right\rangle_{Z}
\left\langle a_r(\mathbf{X},\boldsymbol{Z}^x_{t'})\right\rangle_{Z}
\equiv \left\langle\left\langle  a_r(\mathbf{X},\boldsymbol{Z}^x_{t})a_r(\mathbf{X},\boldsymbol{Z}^x_{t'})\right\rangle\right\rangle_{Z}
\end{gather}
\end{subequations}
The simples assumption for the time dependence of the cumulant $C_{r}^{(2)}$ is exponential decay:
\begin{gather}
\label{Eq::correlation_correction}
C^{(2)}_r(t,t'|\mathbf{X})=K\exp(-\kappa(\mathbf{X})|t-t'|)
\end{gather}
where $\kappa(\mathbf{X})^{-1}$ is a characteristic relaxation time of the regression of fluctuation of species
$\boldsymbol{Z}$ and $K=\langle(\Delta a^2_r(\mathbf{X},Z))\rangle_Z$. In this case correction to the Markovian term is given by:
$$
\Delta \bar{a}_r(t,\mathbf{X})\cong
 -K\kappa^{-1}(\mathbf{X})\frac{\partial}{\partial t}\left(t-\kappa^{-1}(\mathbf{X})[1-\exp(-\kappa({\bf X})t)]\right)
$$
Correction to the Markovian approximation given by
(\ref{Eq::second_order_correction}) is exact for the Gaussian and
Markov process since  the only possible expression for the
correlation function of a stationary Markov and Gaussian process is
the exponential of a form (\ref{Eq::correlation_correction}). It is
also interesting to note that correlation correction
(\ref{Eq::second_order_correction})  generally decreases the transition rate. 
This is a result which can not be obtained using only straightforward averaging method presented
in publications \cite{CVRaoArkin2003,CaoGillespiePetzold2005}.
\par
Note that in general  relations (\ref{Eq::renorm_rates}) can be
viewed as a type of fluctuation-dissipation relations;
 they connect the effective dissipation rate in the slow  coarse-grained dynamics and statistics of fluctuations of
 the fast reaction events given by the cumulants  $C^{(m)}_r(t_1,\ldots , t_m|\mathbf{X})$.
\par

\section{\large Coarse-Grained Dynamics and  Error Control\label{Sec::converge}}
\par
Given the renormalized survival probabilities and  transition
 rates at different points of state space of main species $\bfX$:
$$
\bar{a}_r(t,\bfX)=a_r(\bfX),\quad r\in\mathcal{R}_1
$$
 stochastic dynamics of the main species $X$ can be formulated in the straightforward way, similar to the stochastic simulation algorithm (SSA) \cite{GillespieSSA,DTGillespieBook}.
At the time point  $t=0$ state $\bfX_0$ we consider an overall survival probability:
\begin{gather}
S(t|\bfX_0)=\prod_{r\in \mathcal{R}_{1}\cup\mathcal{R}_3}S_r(t|\bfX_0)
\end{gather}
and define a jump moment of the slow process as a first time $\tau_1$ when $S(t|\bfX_0)$ crosses the value $u$,
 where the last one is a random number  uniformly distributed on the interval $(0,1)$ \cite{MHADavisBook}:
\begin{gather}
\tau_1=\inf\{t>0| S(t|\bfX_0)\leq u\},\quad u\in\mathcal{U}(0,1)
\end{gather}
 Post-jump transition kernel is defined  by the vector of transition probabilities 
\begin{gather}
q_r=\frac{\bar{a}_r(\tau_1,\bfX_0)}{\sum_{r'\in \mathcal{R}_{1,3}}\bar{a}_{r'}(\tau_1,\bfX_0)},
\quad r\in\mathcal{R}_{1,3}
\end{gather} i.e.
reaction event $r^*\in \mathcal{R}_{1,3}$ is selected based on the vector $q_r$ and current state is updated:
$$
 \bfX_{\tau_1}= \bfX_0+\boldsymbol{\nu}_{r^*},\quad t_1=\tau_1,
$$
Then the same procedure is performed starting at the state $\bfX_{\tau_1}$
with generation of the interval $\tau_2$  from the survival probability $S(t|\bfX_{\tau_1})$
 and new state $\bfX_{\tau_1+\tau_2}$ and so on. As a result one obtains a coarse-grained trajectory:
\begin{gather}\label{}
(t_n,\bfX_{t_n}),\quad t_n=\sum_{i=1}^n\tau_i
\end{gather} 
\par
 Question about the overall accuracy and the error control is a delicate question.
Below we  decompose the overall error of the method it into the following main factors:
\begin{enumerate}
\item Error in approximating by coarse grained dynamics:
$$e_1=\sup_{0\leq t\leq T}\mathbb{E}(|\mathbf{X}_{t}-\bar{\mathbf{X}}_{t}|^2)$$ assuming that transition rates
$\bar{a}_r(\cdot)$ can be obtained without error.
\item Approximation and  Monte Carlo error $e_2$ of   $\bar{a}_r(\cdot)$ via the finite number
 of samples representing the dynamics  of $\boldsymbol{Z}_{t}$ at fixed  ${\bf X}$.
\end{enumerate}
Below we discuss step by step leading terms in $e_1,~e_2$.
\par
Estimation of the  error $e_1$  is related to the  answer on the following question:
what possible error is introduced  while  performing  averaging of
rates of reactions in the subsets $\mathcal{R}_{1,3}$ at {\em fixed}
$\mathbf{X}$? \par  It is not hard to see that  this  error is
proportional to the probability of the event that minimal jump time
over the reactions in group $\mathcal{R}_{1}\cup\mathcal{R}_2$ is
smaller then $t$ while the minimal jump-time of reaction in the
group $\mathcal{R}_3$  is larger then $t$:
\begin{gather} \label{Eq::error_averaging}
S_{r}(t|\bfZ)=P\left(\{\min_{r\in\mathcal{R}_3}\tau_{r,2}<t\}\cup\{\min_{r\in\mathcal{R}_{1,3}}\tau_{r}>t\}\right)=
\left\langle\exp(-\int_0^tdt'~a_r(\mathbf{X}_{t'},\boldsymbol{Z}_0))\right\rangle_{X},\quad~r\in\mathcal{R}_3
\end{gather} where average  $\langle\ldots\rangle_X$ is taken
over trajectories $\mathbf{X}^z_{t}$ at {\em fixed} $\mathbf{Z}$
It is not hard to see that this probability is exponentially
small, i.e. $ \propto \exp(-{\epsilon}^{-1}\frac{t}{const})$ in the limit $\epsilon\to 0$.
\par
Error $e_2$ depends on the  number of cumulants we have included in \refe{Eq::cumulant_expansion} and cumulant of order $m$ usually gives contribution proportional to $\epsilon^m$. In Appendix we outline the exact method for calculation
of the renormalized survival probability based on eigenvalue decomposition of certain linear operator which 
is a  practical approach in situations when state space of the variable $\bfZ$ is not very large.
\par

\section{\large Examples\label{Sec::examples}}
\par
\begin{figure}[!h]
\begin{center}
\includegraphics[scale=1.2]{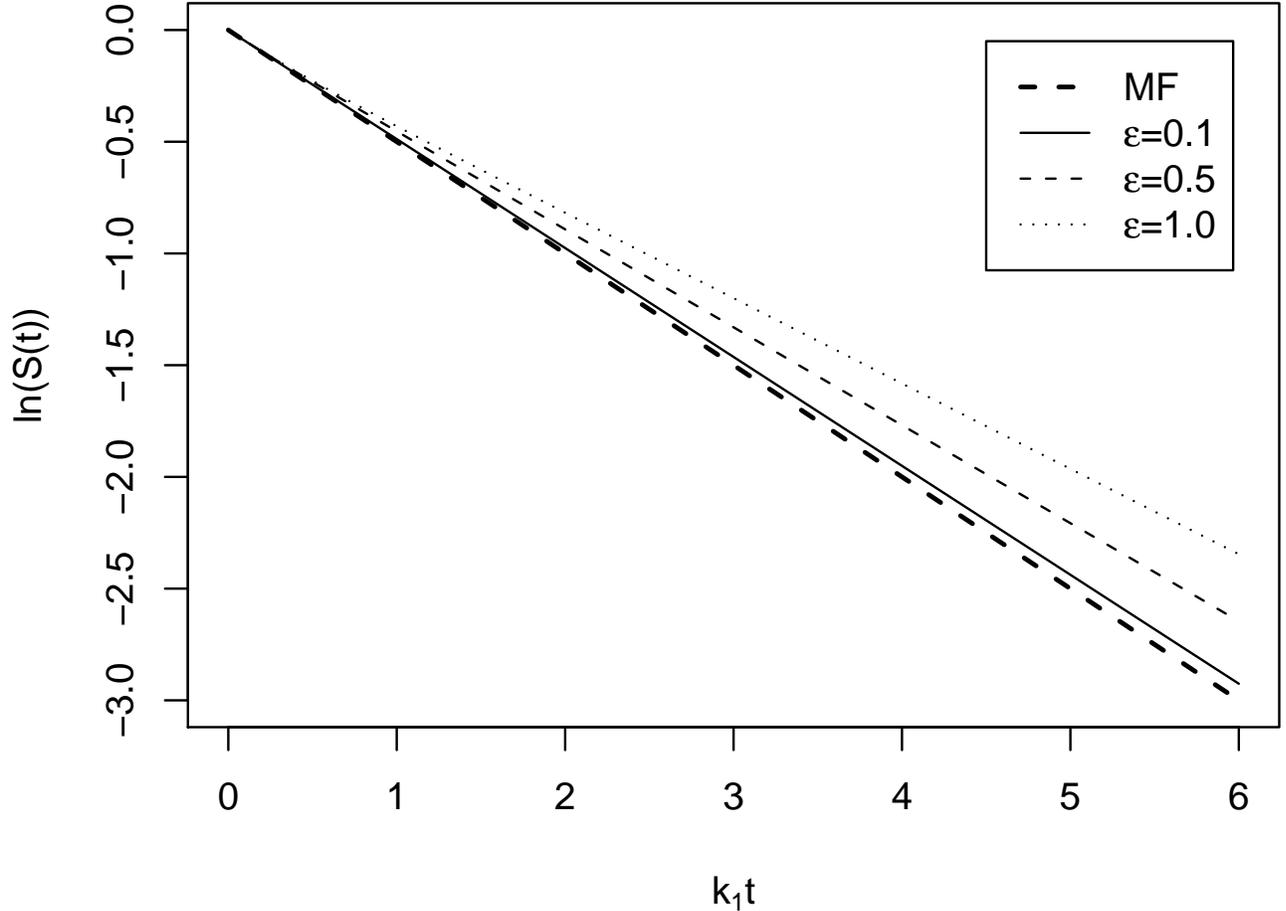}
\caption{ \label{Fig::two_states} 
Time dependence of survival probability $S_r(t)$ for different ratios of transition rates
 $\epsilon={a_r(X,1)/(k_{01}+k_{10})}$ for the system with $a_r(X,1)\neq 0$ and $a_r(X,0)=0$.
}
\end{center}
\end{figure}

  We now present a simple intuitive example to show that exponential or  non-exponential structure of the averaged survival probability  is governed by the relationship between time-scales of  "fast" and "slow" species.
Assume that  for some reaction channel
\begin{gather}
\mathsf{X}+\mathsf{Z}
+\ldots \rightarrow \dots
\end{gather}
 rate $a_r(X,Z)=k_rh_r(X)h'_r(Z)$ jumps reversibly between two values $a_r({X},0)$ and $a_r({X},1)$ 
with the  stochastic dynamics of $Z_{t}$ governed by simple master equation:
\begin{gather}
\begin{pmatrix}
\dot{p}_t(0)\\
\dot{p}_t(1)
\end{pmatrix}=
\begin{pmatrix}
-k_{01} & k_{10}\\
k_{01} & -k_{10}
\end{pmatrix}
\begin{pmatrix}
{p}_t(0)\\
{p}_t(1)
\end{pmatrix}\label{Eq::master_eq_Z_two_levels}
\end{gather} 
\refE{Eq::master_eq_Z_two_levels} describes the switching transitions between the two states $0$ and $1$. 
 Assuming that state of variable $Z$ is prepared  according to the equilibrium density 
$\pi=(\pi_0,\pi_1)=(\frac{k_{10}}{k_{01}+k_{10}},\frac{k_{01}}{k_{01}+k_{10}})$.
the average survival probability $\langle e^{-\int_0^t a_r(X,Z^x_{t'})dt'}\rangle$ can be obtained as follows 
(see  also  Appendix section for the general computational framework):
\begin{gather}\label{Eq::two_level_survival_prob}
S_r(t|\bfX)=
\begin{pmatrix}
1\\
1
\end{pmatrix}^{T}
\exp\left(t
\begin{pmatrix}
-a_r(X,0)-k_{01} & k_{10}\\
k_{01} & -a_r(X,1)-k_{10}
\end{pmatrix}
\right) 
\begin{pmatrix}
\pi_{0}\\
\pi_1
\end{pmatrix}
\end{gather} 
 This result is very similar in nature to the result obtained in \cite{AgmonHopfield83} for the case of identical transition rates. Remarkable and simple result outlined by \refe{Eq::two_level_survival_prob} allows us to capture in essence  regimes corresponding to the different ratios of the time-scales:
 $a_r\ll (k_{10}+k_{01})$  and $a_r\geq (k_{10}+k_{01})$.
First regime ($a_r\ll (k_{10}+k_{01})$) corresponds to the situation when transitions between 
different states of $Z$ happens much faster then the average rate $a_r(X,0),a_r(X,1)$ of the "linker" process  
 and represents the  mean-field (MF) regime. In this case dependence of  $ln(S_r(t))$ on time $t$  can be very well characterized as linear \reff{Fig::two_states}.  Not surprisingly, other regime, i.e. $a_r\gg (k_{10}+k_{01})$ can be characterized as gated:
in this case effective transition rate $\bar{a}_r$ is characterized by the rate of switching of $Z$: $k_{01}+k_{10}$.
\par
\begin{figure}[!h]
\begin{center}
{
\includegraphics[scale=1.2]{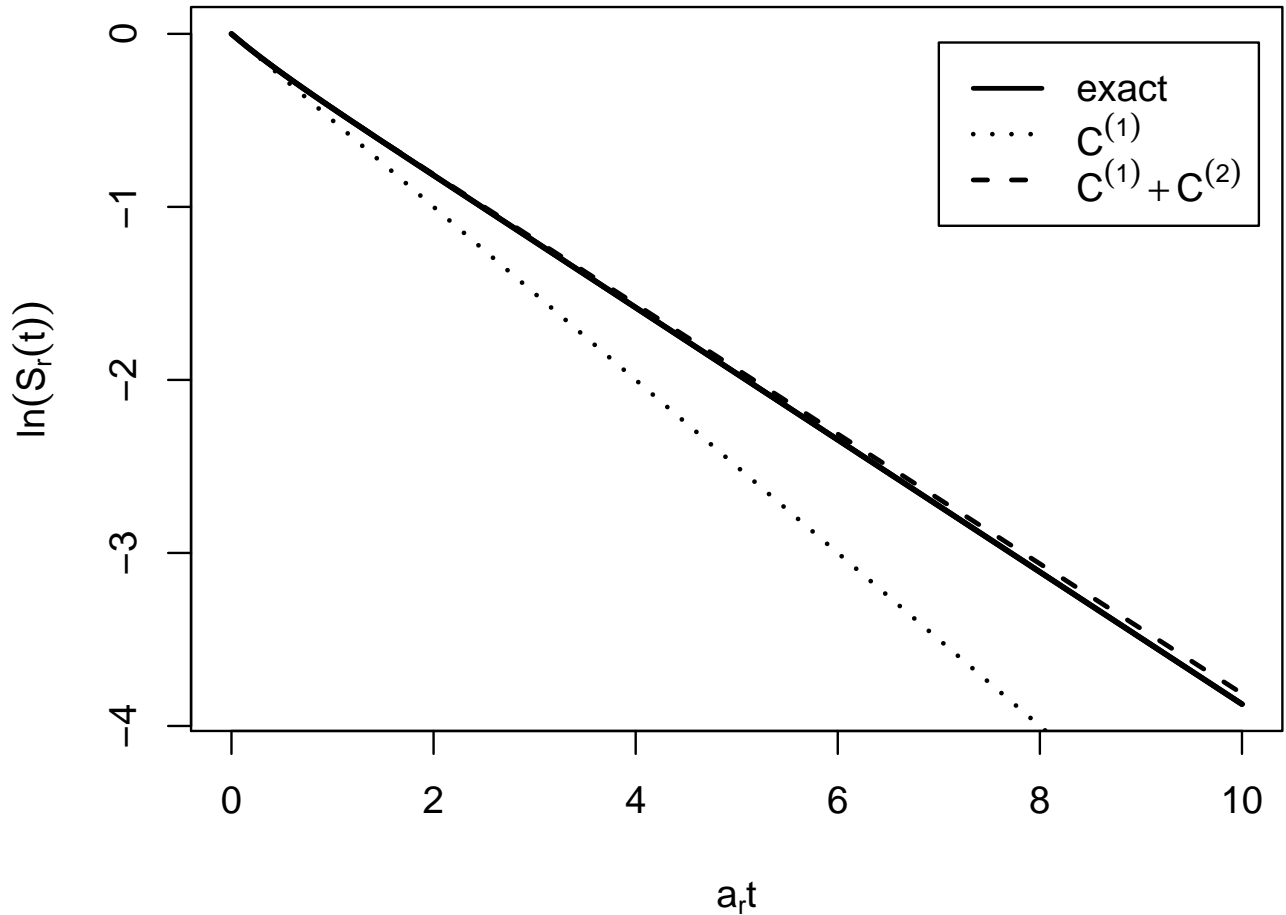}
}
\caption{\label{Fig::two_states_cumulant} Time dependence of the survival probability $S_r(t)$ calculated 
with mean-filed (dotted line) approximation and second cumulant correction (dashed line) compared to exact dependence
(solid line).}
\end{center}
\end{figure}

Figure \ref{Fig::two_states_cumulant} demonstrates  influence of the second order correlation correction \refe{Eq::correlation_correction}:
$\Delta \bar{a}_r(t,\bfX)= \pi_1\pi_0 \frac{t}{\kappa}(1-\frac{\kappa}{t}(1-e^{-\kappa t})),~\kappa=k_{01}+k_{10}$
which fluctuation correction to the effective rate $\bar{a}_r(\cdot)$
\par
Dependence of survival probability $S_r(t|\bfX)$ in the example of a two-state system  can be shown to be  non-exponential on the longer time scale but  $\ln(S_r(t))$ behaves linearly with time at small times $t\leq 1/a_r(X,\cdot)$. 
\par
\begin{figure}[!h]
\begin{center}
{
\includegraphics[scale=0.8]{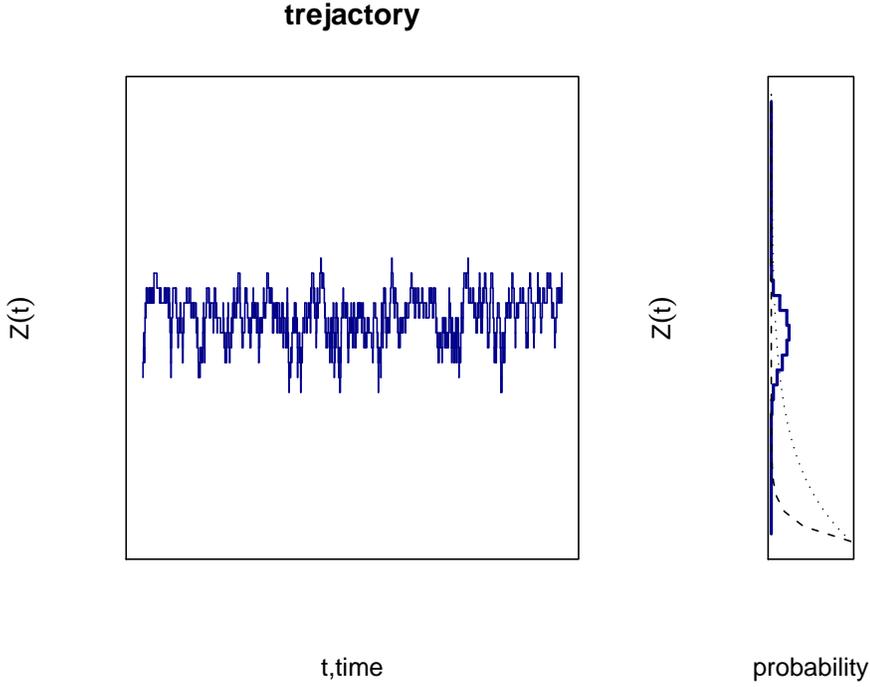}
}
\caption{\label{Fig::fast_process_plot} Trajectory and probability density of the process $Z(t)$. Dotted and dash lines
on the probability plot correspond to the profile of the relaxation rate $a_r(X,Z)$ for different $X$. }
\end{center}
\end{figure}
\begin{figure}[!h]
\begin{center}
{
\includegraphics[scale=1.2]{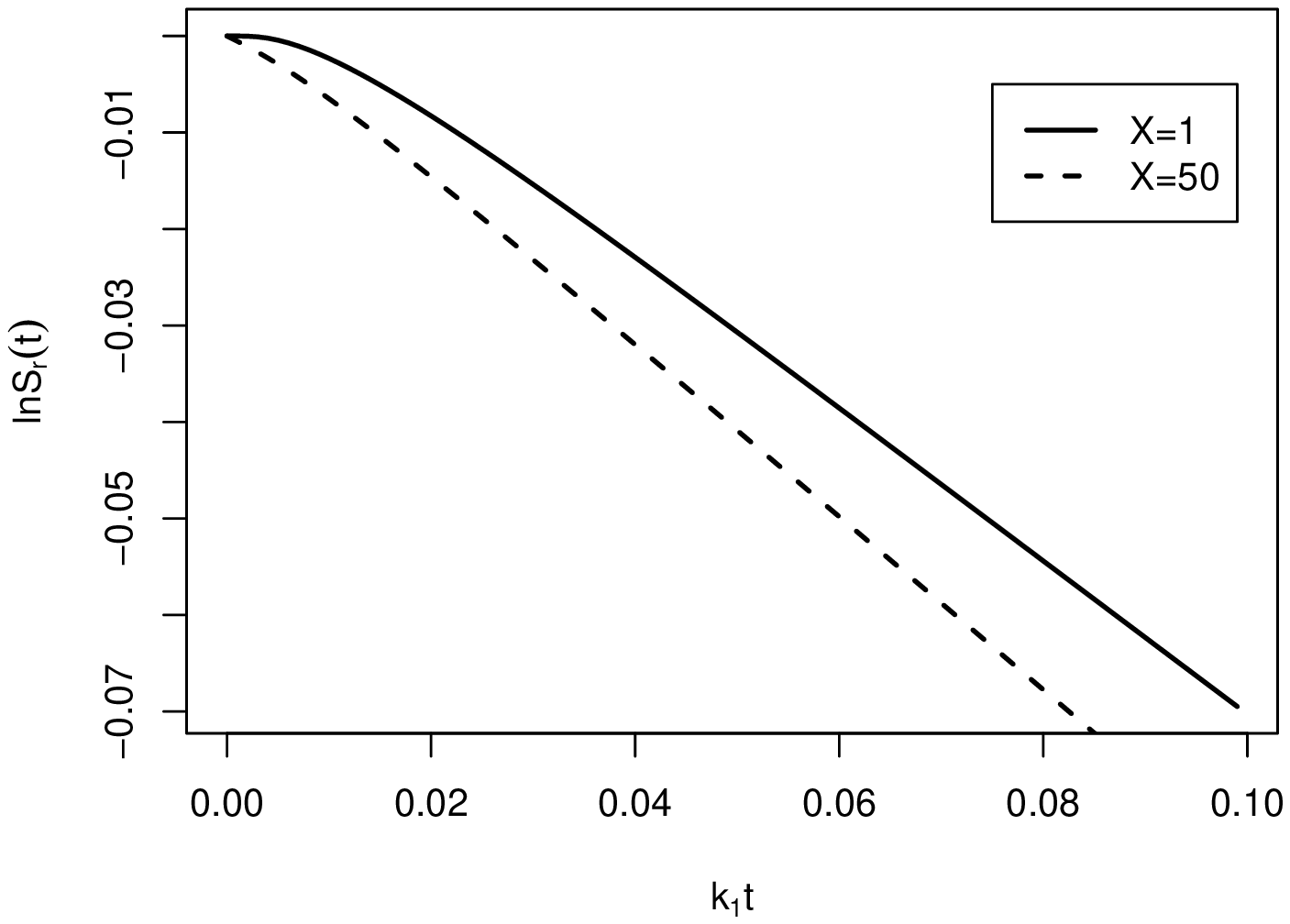}
}
\caption{\label{Fig::non_exponential}
Time dependence of the survival probability $S_r(t|X)$ for the system 
where dimerization dynamics of the fast variable $Z$ is described by parameters 
$N_m=200,~k_1=1.0,~k_2=10.0,~K_{eq}=10^2$. Plots are shown for values of $X=1$ and $50$ clearly manifest
non-exponential character of the relaxation process at small time for low values of $X$. Note that kinetics is non-exponential on time 
larger then characteristic scale $t_{non-exp}\approx 0.02$ of fluctuation of $Z$ ( $k_1^{-1}(N/2)^2\approx 10^{-3}$) i.e. on the relevant for time-coarsening interval.
}
\end{center}
\end{figure}

\vskip 1cm
Interesting case of non-exponential relaxation kinetics, and specifically non-exponential kinetics at small times 
 can be presented by the following example. Consider a fast reaction given  by the dimerization reaction:
\begin{gather}
\mathsf{S}+\mathsf{S}\overset{k_2}{\underset{k_2K_{eq}}{\rightleftarrows}}  \mathsf{S}_2
\end{gather} 
\begin{subequations}
where the fast variable $Z_{t}$ is the number of reaction event which took place up to time $t$ which relates 
the numbers of monomers and dimers with the total number of molecules $N_m=2S+S_2$ in the following way:
\begin{gather}
S=N_m-2Z,\quad
S_2=Z
\end{gather}
\end{subequations}
and a "linker" process is  described by the relaxation rate depending on the number of dimers $X$ in the following way:
\begin{gather} 
a_r(X,Z)=\frac{k_1X}{Z+X}\label{Eq::relax_rate_nonexp}
\end{gather}
Current value $X$ serves as an activation threshold: 
at small values of $X$ ($X\propto 1$) only  small values of $Z$ contribute to the effective rate but probability  that $Z$ takes values away from its average are exponentially suppressed (\reff{Fig::fast_process_plot}). 
On the contrary, if $X$ is large i.e. $X \approx \sum_{Z} \pi(Z|X) Z$ then rate given by \refe{Eq::relax_rate_nonexp}
depends  on the typical  value of $Z$ and $S_r(t|X)$ manifests time dependence similar to the previous example.
One can see that this relaxation process shows non-exponential time dependence at small times
due to the fact that process $Z_{t}$  rarely visits the states contributing to the maximum of the relaxation rate given by 
\refe{Eq::relax_rate_nonexp}. 
We investigate the dependence of the averages survival probability on the level of activation threshold $X$ and 
value of the equilibrium constant $K_{eq}$. Results presented on the  \reff{Fig::non_exponential}  show non-exponential behavior of averaged survival probability  for the system  at small times $t$. It is evident that non-exponential behavior of $S_r(t|X)$ is less pronounced for large values of $X (~X\approx \langle Z\rangle)$.
\par
Eigenvalue-eigenvector decomposition and calculation of expansion coefficients was performed via standard routines of
 LAPACK library available at {\small\tt http://www.netlib.org }.\par
\section{\large Discussion and Conclusions}
\par
Let us summarize the main aspects of this paper. We have studied 
 reduction approach to eliminate a fast intermediate in the chemical reaction network.
To develop this method it is important to consider the time coarse-grained transition rates.  We have discussed the limitations of the principle of stochastic averaging and its possible extensions through the rigorous technique for construction of the effective transition rates. We outline the procedure for re-normalization of the transition rates and construction of the effective Markov chain for the slow reactions. The merit of the present approach is that it is based on a conceptually transparent probabilistic approach involving the waiting-time distribution.Technique itself 
resembles a non-Markovian generalization of the Kubo-Anderson theory of stochastic modulation.
Our study clearly indicates importance of details of the statistical structure of averaging process.
\par
\section{Acknowledgments}
\par
Author thanks  A. Alekseyenko for stimulating discussion on the
subject  of this publication, T. Ham for valuable suggestions.
Author would like to acknowledge DARPA grant \# BAA-01-26-0126517 and Prof. A.P.Arkin 
for support during the course of this research. 
\bibliographystyle{unsrt}
\bibliography{multi_scale}

\pagebreak\newpage
\par
\section{\large Figures}
\begin{figure}[!h]
\begin{center}
\includegraphics[scale=0.4]{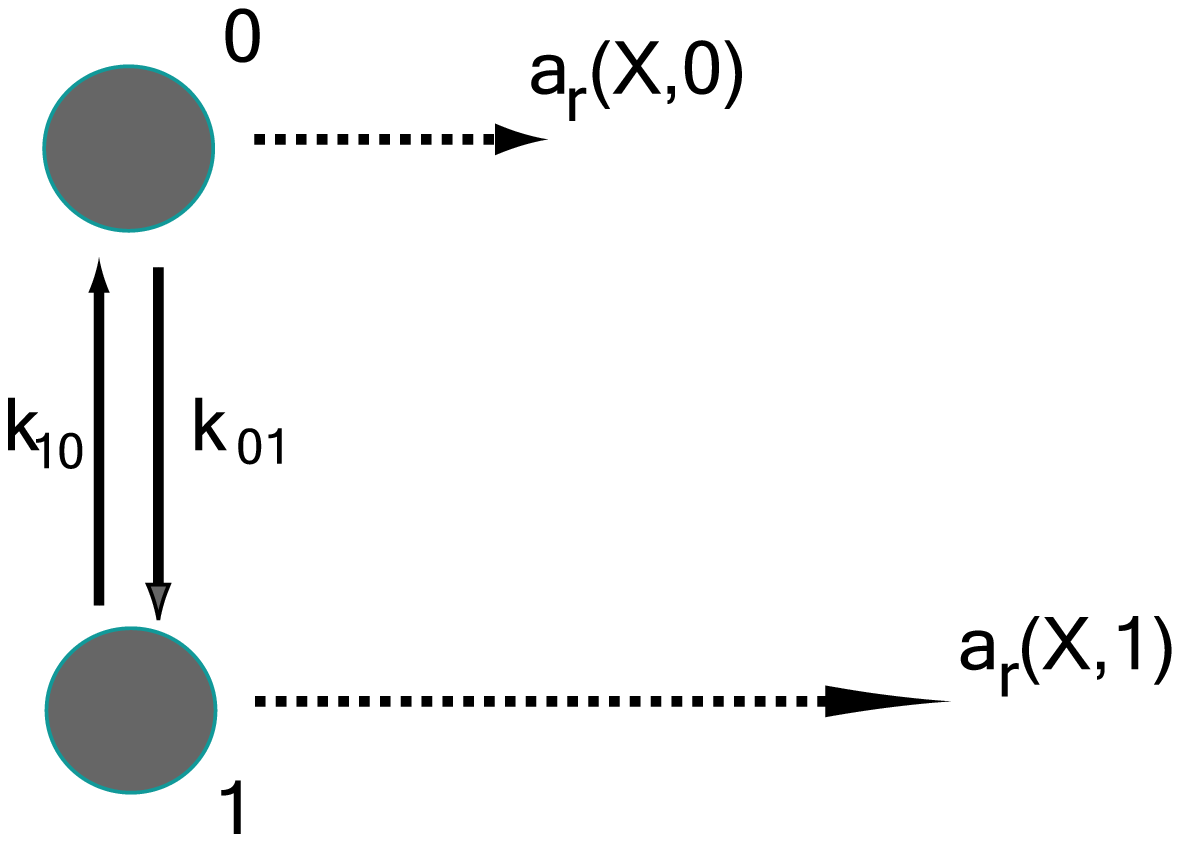}
\caption{\label{Fig:two_level} Schematic representation of the two-state model. Relaxation rates $a_r(\cdot)$
depend on both state $Z$ and $X$ and can be quite general.}
\end{center}
\end{figure}

\pagebreak\newpage

\appendix{
\section{Calculation of averaged survival probability \label{Appendix}}
\par
Calculations of averaged survival probabilities $S_r(t|\bfX)$ requires, in general,
 the calculation of the cumulants $C_r^{(m)}$ of  different order $m$  but for some simple cases it 
can be obtained exactly. 
This is possible for the class of systems which have only finite number of accessible states of the fast variables. \par
One can study the distribution of  values $S$ of the functional 
\begin{gather}\label{Eq:exponential_functional}
\exp(-\int_0^t a_r(\boldsymbol{Z}_{t'})dt'),
\end{gather} where we have omitted the current state $\mathbf{X}$ to simplify the notation.
We introducing the joint probability density $q(S,\boldsymbol{Z},t)$
of the  random variables $S$ and $\boldsymbol{Z}$ \cite{RiskenBook}:
\begin{subequations}
\begin{gather}
\frac{\partial q(S,\boldsymbol{Z},t)}{\partial t}=
a_r(\boldsymbol{Z})\frac{\partial }{\partial S}(Sq(S,\boldsymbol{Z},t) )+\\
+\sum_{r'\in \mathcal{R}_2} \left(a_{r'}(\boldsymbol{Z}-\boldsymbol{\nu}_{r'}))
q(S,\boldsymbol{Z}-\boldsymbol{\nu}_{r'},t)-a_{r'}(\boldsymbol{Z}))
q(S,\boldsymbol{Z},t)\right)=\\=
a_r(\boldsymbol{Z})\frac{\partial }{\partial S}(Sq(S,\boldsymbol{Z},t) )
+\sum_{Z'}\mathbb{W}_{\boldsymbol{ZZ'}}q(S,\boldsymbol{Z},t)
\end{gather}

Average survival probability can be expressed following:
\begin{gather}
S_{r}(t)=\sum_{\boldsymbol{Z}}\int_0^1 S q(S,\boldsymbol{Z},t) dS=
\sum_{\bfZ}\bar{q}_r(\bfZ,t)
\end{gather}
and $\bar{q}_r(Z,t)$ is governed by  the following master equation:
\begin{gather}
\frac{\partial \bar{q}_r(\boldsymbol{Z},t) }{\partial t}= -a_r(\boldsymbol{Z})\bar{q}_r(\bfZ,t) +\sum_{\boldsymbol{Z'}}
\mathbb{W}_{\boldsymbol{ZZ'}}\bar{q}_r(\boldsymbol{Z}',t)
\end{gather}
\end{subequations}
One can find an averaged survival probability via eigenvalue-eigenvector $\{ \lambda,  V_\lambda(\boldsymbol{Z})\}$
 decomposition of the linear operator 
$\mathbb{W}_{\boldsymbol{ZZ'}}-a_r(\boldsymbol{Z})\delta_{\bfZ\bfZ'}$:
\begin{gather}
S_r(t)=\sum_{Z}\sum_{\lambda}c_{\lambda}V_\lambda(\boldsymbol{Z})\exp(\lambda t)
\end{gather}
where coefficients $c_\lambda$ correspond to the decomposition of the invariant  probability $\pi(\boldsymbol{Z}|\cdot)$:
\begin{gather}
\pi(\boldsymbol{Z})=\sum_{\lambda}c_{\lambda}V_\lambda(\boldsymbol{Z})
\end{gather}

}
\end{document}